\documentclass[preprint2]{aastex}
\shorttitle{tt}
\shortauthors{at}

\begin{document}

\title{On the shapes and structures of high-redshift compact galaxies}

\author{M\'elanie Chevance\altaffilmark{1,2,3}, Anne-Marie Weijmans\altaffilmark{2}, Ivana Damjanov\altaffilmark{4,1}, Roberto G. Abraham\altaffilmark{1}, Luc Simard\altaffilmark{5}, Sidney van den Bergh\altaffilmark{5}, Evelyn Caris\altaffilmark{6} \& Karl Glazebrook\altaffilmark{6}
}
\altaffiltext{1}{Department of Astronomy and Astrophysics, University of Toronto, 50 St. George Street, Toronto, ON, M5S 3H4, Canada}
\altaffiltext{2}{Dunlap Institute for Astronomy \& Astrophysics, University of Toronto, 50 St. George Street, Toronto, ON, M5S 3H4, Canada}
\altaffiltext{3}{D\'epartement de Physique, \'Ecole Normale Sup\'erieure de Cachan, 61 Avenue du Pr\'esident Wilson, 94235 Cachan Cedex, France}
\altaffiltext{4}{Harvard-Smithsonian Center for Astrophysics, 60 Garden Street, MS-20, Cambridge, MA 02138, USA}
\altaffiltext{5}{Dominion Astrophysical Observatory, Herzberg Institute of Astrophysics, National Research Council of Canada, 5071 W. Saanich Road, Victoria, BC, V9E 2E7, Canada}
\altaffiltext{6}{Centre for Astrophysics and Supercomputing, Swinburne University, Hawthorn VIC 3122, Australia}


\begin{abstract}
Recent deep {\it Hubble Space Telescope} WFC3 imaging suggests that a majority of compact quiescent massive galaxies at $z \sim 2$ may contain disks. To investigate this claim,  we have compared the ellipticity distribution of 31 carefully selected high-redshift massive quiescent compact galaxies to a set of mass-selected ellipticity and S\'ersic index distributions obtained from 2D structural fits to $\sim$$40,000$ nearby galaxies from the Sloan Digital Sky Survey. A Kolmogorov-Smirnov test shows that the distribution of ellipticities for the high-redshift galaxies is consistent with the ellipticity distribution of a similarly chosen sample of massive early-type galaxies. However the distribution of S\'ersic indices for the high-redshift sample is  inconsistent with that of local early-type galaxies, and instead resembles that of local disk-dominated populations. The mismatch between the properties of high-redshift compact galaxies and those of both local early-type and disk-dominated systems leads us to conclude that   the basic structures of high-redshift compact galaxies probably do not closely resemble those of any single local galaxy population. Any galaxy population analog to the high-redshift compact galaxies that exists at the current epoch is either a mix of different types of galaxies, or possibly a unique class of objects on their own. \\
\end{abstract}

\section{Introduction}

With the discovery of very compact massive quiescent galaxies at redshifts $z\sim2$ \citep[e.g.][and references therein]{Daddi2005,Toft2007, Trujillo2007, Zirm2007,Buitrago2008,Cimatti2008,VanDokkum2008,Damjanov2009,Williams2010,Saracco2010,Szomoru2011}, galaxy formation and evolution models have been challenged to explain the increase in size of these systems with decreasing redshift and the fact that similarly compact massive galaxies are almost completely absent from the local Universe \citep{Trujillo2009,Taylor2010,Valentinuzzi2010a}. Compared to present-day galaxies of similar (stellar) mass ($M_* \simeq 10^{11}M_\odot$), these high-$z$ compact galaxies or 'red nuggets' are a factor of $\sim5$ smaller \citep[e.g.,][]{Damjanov2011} with half-light or effective radii $R_e\lesssim1$ kpc, 
while their stellar mass densities within the effective radius are on average an order of magnitude higher \citep[although this difference is smaller for the stellar mass densities within the central kiloparsec, see e.g.][]{Bezanson2009,Saracco2012}. Based on the low fraction of close pairs among quiescent galaxies observed at $0<z<2$ \citep{Bell2006a, Bundy2009, DePropris2010, Williams2011,Man2012} and the small number of near equal mass mergers produced in N-body simulations \citep[e.g.,][]{Shankar2010}, it seems that major mergers can only be partly responsible for the observed size growth. Additional secular processes, such as adiabatic expansion \citep{Fan2010, Damjanov2009} and/or a series of minor mergers \citep{Naab2009, Hopkins2010,Oser2012}, might be needed to expand these compact systems. 

Recent deep, high-resolution images taken with the Wide Field Camera (WFC3) and NIC2 camera on board the {\it Hubble Space Telescope} (HST) have shown that based on observed elliptiticies and S\'ersic profile fits, there may be indications that many of these compact high-redshift galaxies contain disks \citep{VanDokkum2008,VanderWel2011}. Indeed, \citet{VanderWel2011} claim based on their sample of 14 quiescent massive compact galaxies, that the majority of these systems are disk dominated. Based on the structural parameters derived from ground-based imaging, \citet{Whitaker2012} have recently reported a small decrease in the average axial ratio with redshift (i.e., mildly increased prominence of the disk component) for a sample of post-starburst galaxies found at $0<z<2$. These findings would have interesting consequences for the possible formation and evolution scenarios of these galaxies \citep[e.g.,][]{Weinzirl2011}. In the present paper we investigate the claim in \citet{VanderWel2011} by comparing the ellipticity and S\'ersic distributions of these objects to the corresponding distributions for nearby massive disk-dominated galaxies, as well as massive bulge-dominated early-type galaxies. To define our high-redshift sample we complement the \citet{VanderWel2011} sample with the \citet{Damjanov2011} synthesis of published structural data for high-$z$ compact galaxies with secure spectroscopic redshifts. At low redshift, we characterize the ellipticity and S\'ersic index distributions of nearby galaxies using the local morphology sample recently published by \citet{Simard2011}.

\section{Data}

Our sample of high-redshift massive compact galaxies consists of 14 galaxies with photometric redshifts $z>1.5$ taken from \citet{VanderWel2011}, augmented by objects selected from the compendium published by \citet{Damjanov2011}. All 69 galaxies in the Damjanov et al. sample have spectroscopic redshifts, HST-based images and published ellipticities and S\'ersic indices. Around 1/4 of these galaxies are at $z>1.5$, have stellar masses $M_*>10^{10.8} M_\odot$, and are compact ($R_e < 2$ kpc)  and quiescent (not actively forming stars). These 17 objects are added to our high-redshift sample, resulting in a total sample size of 31 galaxies. Galaxies in the Damjanov et al. sample span a mass range from $10^{9.8}M_{\sun}$ to $10^{12.1}M_{\sun}$, with a median of $10^{11.1}M_{\sun}$. Galaxies in the \citet{VanderWel2011} sample all have stellar masses larger than $10^{11.1}M_{\sun}$, with a median of $10^{11.3}M_{\sun}$ when scaled to a Salpeter (1955)  initial mass function to be consistent with the Damjanov et al. sample. The two samples therefore include objects with similar stellar masses and in the same redshift range, and all have $R_e < 2$ kpc.

Our local galaxy sample is selected from the morphological catalog recently published by \cite{Simard2011}. This catalog
is based on two-dimensional structural fits (S\'{e}rsic bulge$+$ exponential disk decompositions, as well as single S\'ersic fits) in the {\itshape g} and {\itshape r} bandpasses  for 1.12 million local galaxies taken from the Sloan Digital Sky Survey.
We restrict our comparison to galaxies with redshifts $z<0.1$ in order to minimize the impact
of seeing on the observed shapes. As described in \citet{Simard2009}, a robust sample of early-type galaxies can be
selected from this parent catalog by requiring that galaxies have
a bulge fraction $B/T>0.5$ and an image smoothness $S2\leq0.075$\footnote{As described below, we experimented with changing these parameters to select even cleaner subsets of early-type galaxies. All of the results in the present paper are robust to modest changes in these parameters.}. Imposing these cuts results in a sample of $\sim 127,000$ early-type galaxies, which defines our local early-type galaxy parent sample. Our local disk-dominated sample was chosen solely based on bulge-to-total ratio, with $B/T <0.5$. Stellar masses, $M_{*}$, were estimated using the approximation
 presented by \cite{Bell2003}:
$\log(M_{*}/M_{\sun}) = -0.499 + 1.519 (M_{g} - M_{r})- 0.4\times(M_{g} - M_{g\sun}) - \log(0.7)$
where $M_g$ and $M_r$ are  rest-frame absolute magnitudes in {\itshape g-}band and {\itshape r-}band, respectively, and $M_{g\sun}=5.12$ is the g-band absolute magnitude of the sun. A \citet{Salpeter1955} initial mass function (IMF) was assumed in deriving this relation. The relative masses determined using this formula are likely to be accurate to 0.1 dex over the redshift range of our local sample \citep{Bell2003}.
Fair comparisons between the high-redshift and low-redshift samples require careful matching of the mass ranges of the samples, and most of the comparisons in this paper will be based on mass-selected subsets of our local galaxy sample. Note that our selection of our local samples is based on morphology only, and does not put any restrictions on star formation activity or quiescence.

\section{Analysis}

\begin{figure*}
\begin{center}
\includegraphics[width=8cm]{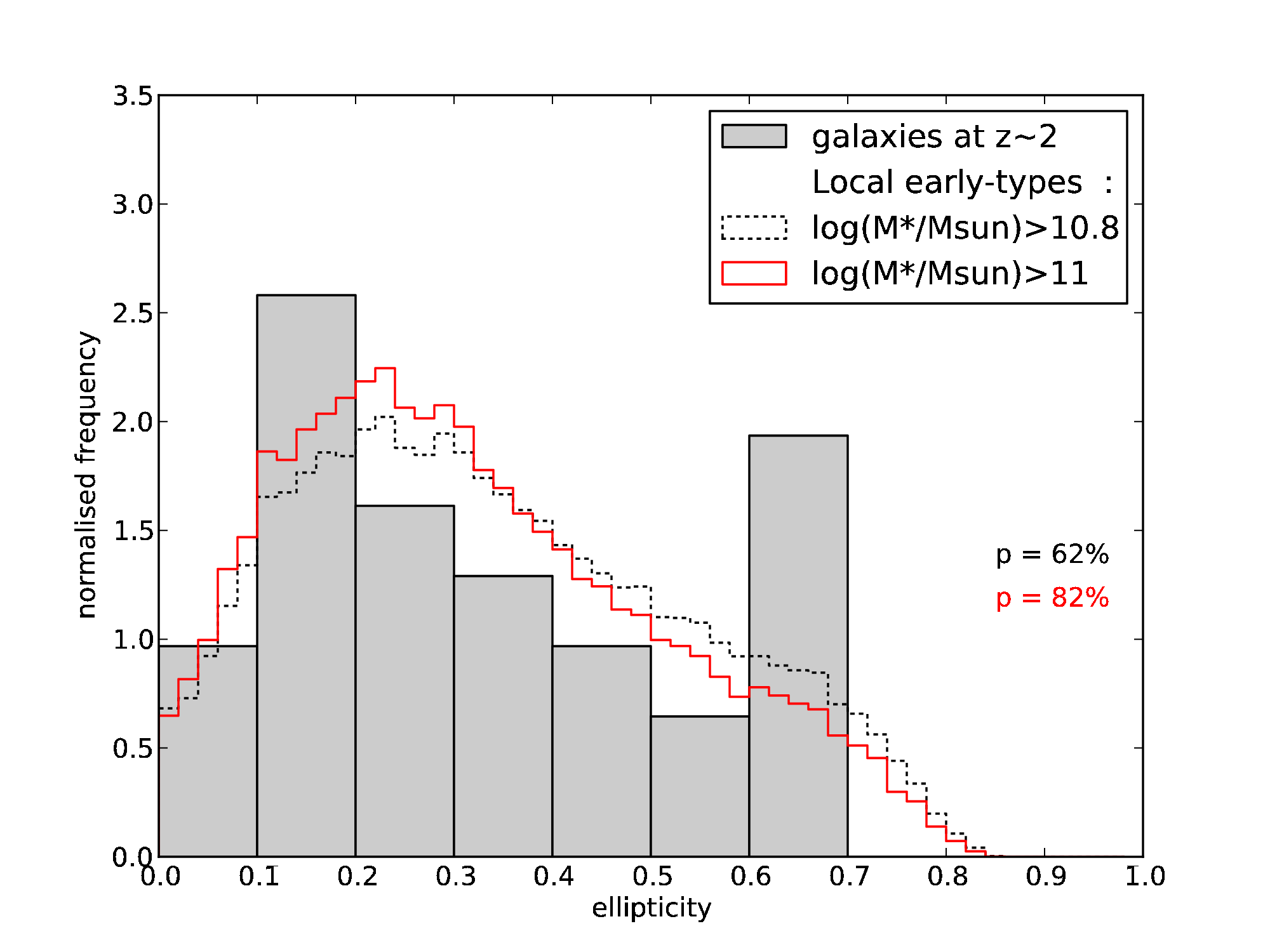}
\includegraphics[width=8cm]{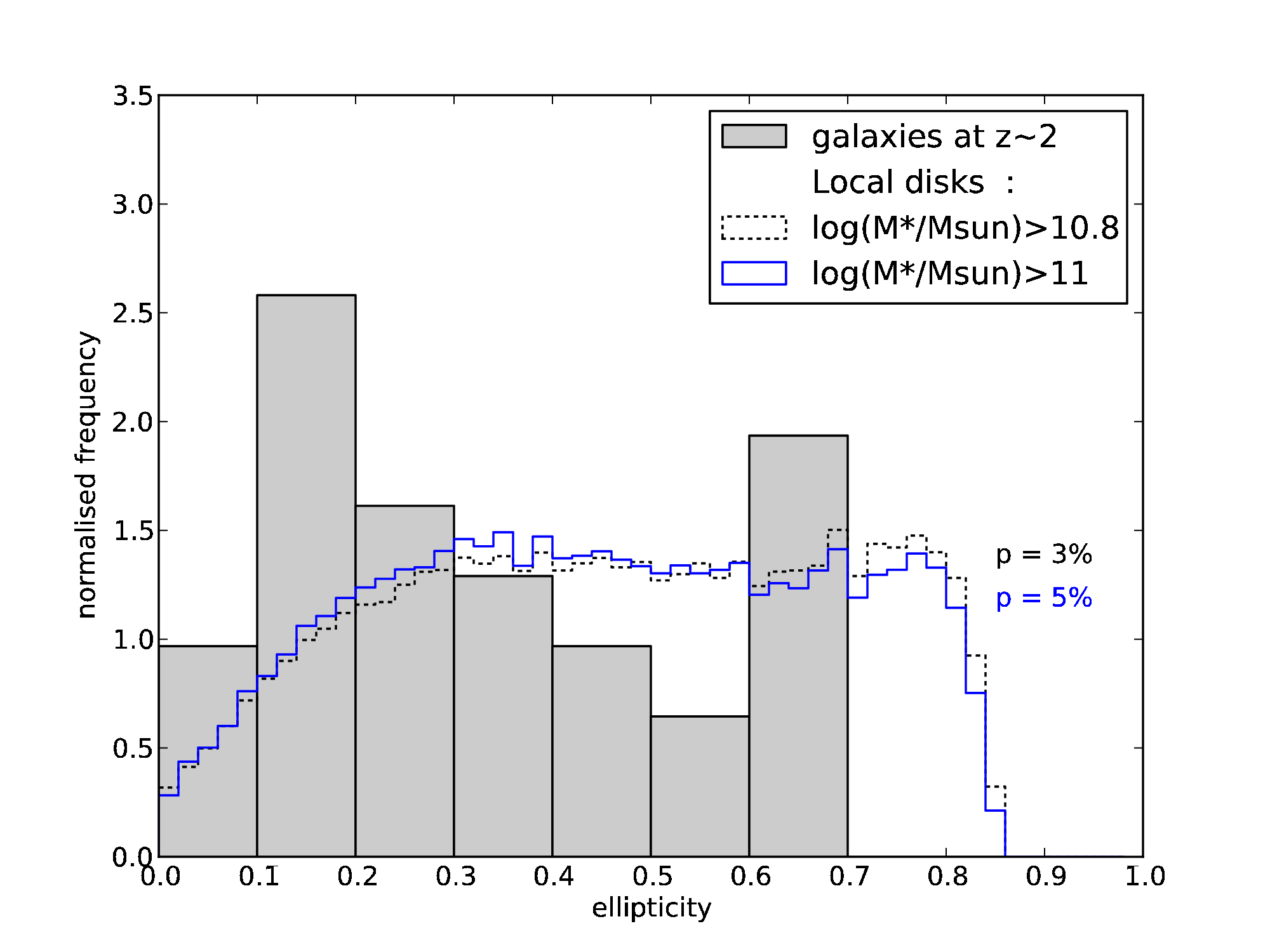}
\caption{(Left:) The ellipticity distribution for 14 compact massive galaxies at $z\sim2$ from van der Wel et al. (2011), complemented by 17 high-redshift massive compact galaxies from Damjanov et al. (2011) (broad grey bins). These data are compared to the distributions of local massive early-type galaxies at $z<0.1$ from Simard et al. (2011). Solid red and dotted black lines show the ellipticity distributions of local early-type galaxies with
 $M_{*}>10^{11} M_{\odot}$ and  $M_{*}>10^{10.8} M_{\odot}$, respectivly.
The labeled percentages are the corresponding $p$-values from a Kolmogorov-Smirnov test which gives the probability that the high-redshift data
and low-redshift data originate from the same distribution. (Right:) As for the panel at left, but now comparing the ellipticitity distribution of the high-redshift massive compact galaxies from van der Wel et al. (2011) and Damjanov et al. (2011) to the distribution of massive nearby disk-dominated galaxies at $z<0.1$.}
\label{fig1}
\end{center}
\end{figure*}

Figure \ref{fig1} presents the apparent ellipticity distributions for our high-redshift massive compact galaxy sample. The ellipticity is as usual defined as $\epsilon = 1 - b/a$, with $b/a$ the observed axis ratio of the images, such that round galaxies have $\epsilon = 0$. These ellipticities (and also the S\'ersic indices that we introduce below) have been measured on the total galaxy image using the two-dimensional fitting routine GIM2D, so prior to bulge-disk decomposition \citep{Simard2011}. In the left-hand panel, ellipticity distributions for our local early-type galaxy samples have been superposed on the high-redshift distribution.  The local early-type galaxies samples consist of objects with stellar masses $>10^{10.8}M_{\odot}$ (solid red line) and early-type galaxies with stellar masses $>10^{11}M_{\odot}$ (dotted purple line). The $>10^{11}~M_{\odot}$ curve provides the most meaningful comparison to the compact massive galaxies sample,  but both local galaxy distributions are shown in order to illustrate the robustness of our conclusions to small (0.2 dex) variations in the mass threshold. A Kolmogorov-Smirnov (K-S) test was used to determine whether the low-redshift distributions differ significantly from the high-redshift distribution, and the corresponding $p$-values from the test are shown as labels in the figure. The high-redshift and local early-type galaxy ellipticity distributions differ at a less than 1$\sigma$ level and are therefore statistically indistinguishable.
Our conclusion remains the same if we choose a more conservative cut ($B/T>0.7$) in defining our local early-type galaxy sample ($p=77\%$ and $p=67\%$ for $M_*>10^{10.8} M_\odot$ and $M_*>10^{11} M_\odot$  respectively). 

Although the high and low-redshift ellipticity distributions presented in the left panel of Figure~\ref{fig1} appear consistent, the small number of galaxies included in the high-$z$ sample raises the question whether there are any distributions at all that could be ruled out using such a small sample. It is thus rather interesting to show
that one distribution that can clearly be eliminated is that of  local disk-dominated galaxies. The right-hand plot in Figure~\ref{fig1} shows the results of K-S tests comparing our sample of 31 high-redshift massive compact galaxies to the ellipticity distributions of similarly massive disk-dominated ($B/T<0.5$) galaxies from our local sample. The ellipticity distribution for this local population decreases for $\epsilon <0.2$, indicative of earlier findings that local disks are not intrinsically circular \citep[e.g.][]{Ryden2004}. There is only a $5\%$ probability that the high and low-redshift ellipticity distributions are drawn from the same population. For a more conservative cut ($B/T < 0.35$), this probability drops to $2\%$. Therefore, on the basis of ellipticity distributions alone, one is driven to conclude that the population of compact objects is dominated by early-type galaxies. Our conclusion remains unchanged if we select from our local samples only those galaxies that have effective widths (EW) of the [OII] line less than 5\AA, indicating a quiescent (non-star forming) nature. For local massive ($M_* > 10^{11}M_{\odot}$) quiescent early-type galaxies, $p=82\%$, while for the local massive quiescent disk galaxies, the probability that they come from the same distribution as the high-redshift compacts stays low at $p=11\%$. However, is the same conclusion reached when other structural data are also considered? 

\begin{figure*}[tb]
\begin{center}
\begin{tabular}{cc}
\includegraphics[width=8.0cm]{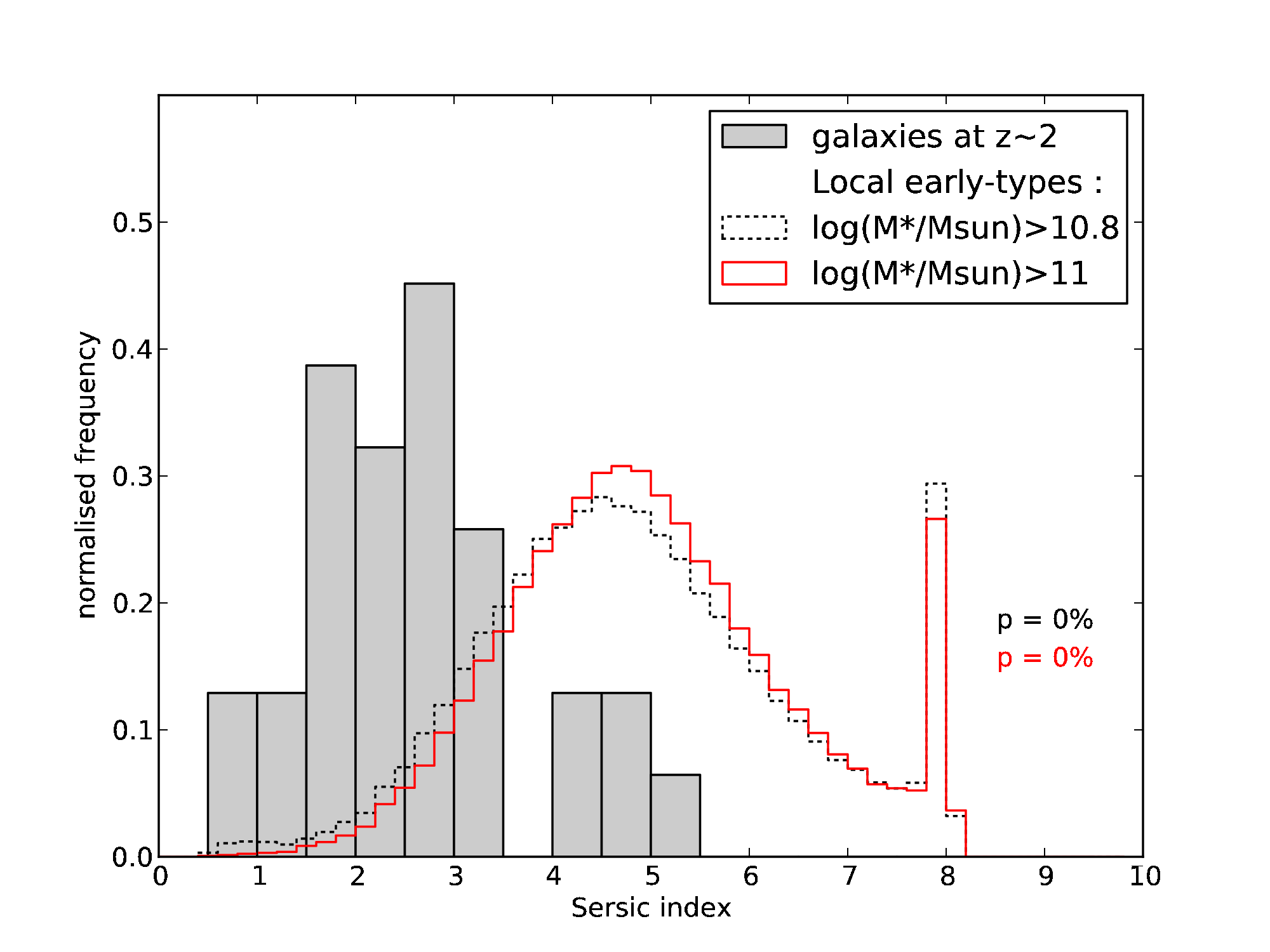} &
\includegraphics[width=8.0cm]{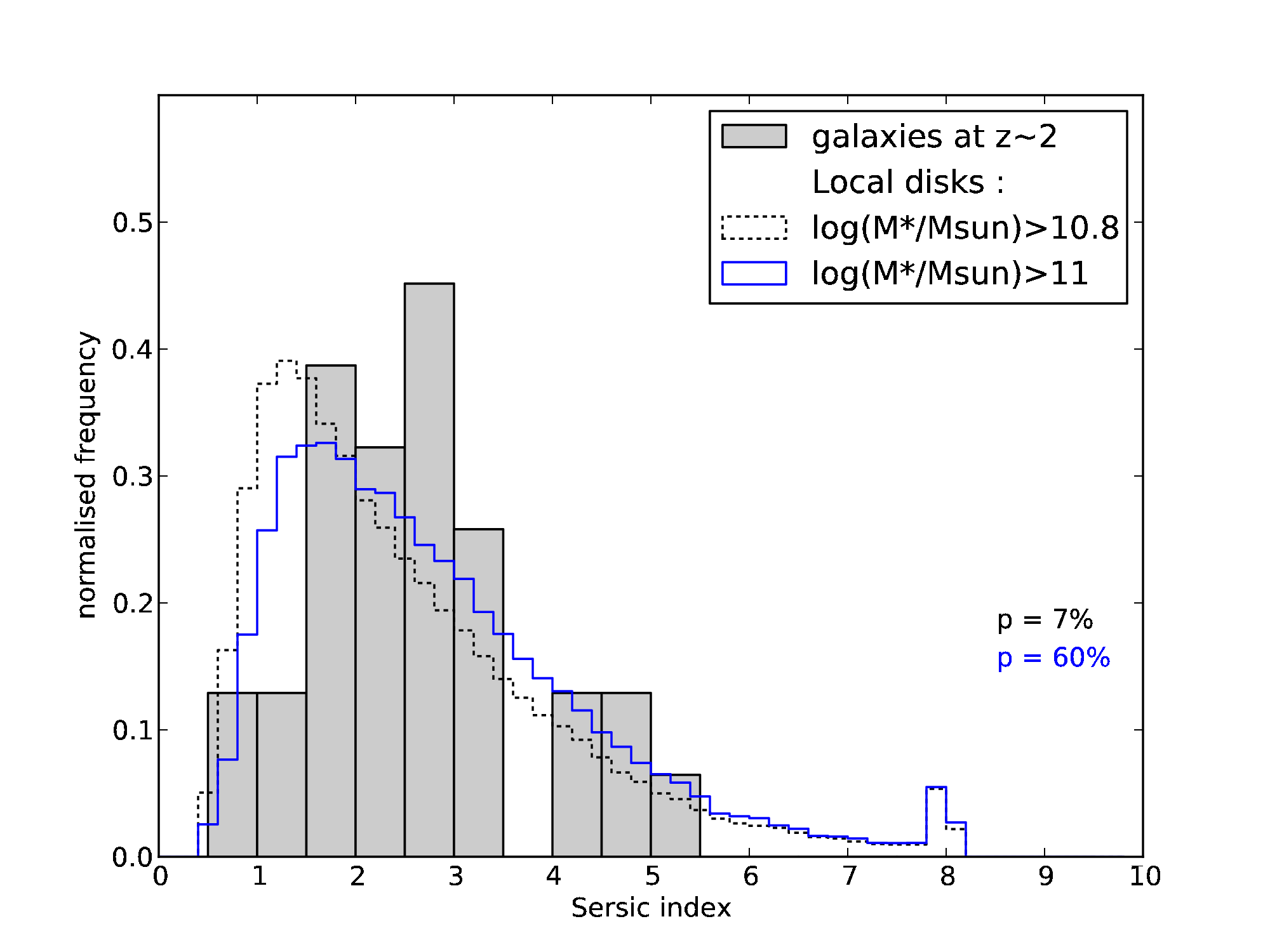} 
\end{tabular}
\end{center}
\caption{As Figure~\ref{fig1}, but now comparing S\'ersic index distributions. The excess of objects with $n \gtrsim 7$ consists mostly of galaxies with  a nuclear source or a bar, or a combination thereof (Simard et al. 2011).}
\label{fig_sersic}
\end{figure*}

A simple structural parameter that nicely complements ellipticity is the S\'ersic index, $n$, which is essentially
a measure of galaxy compactness. This parameter is strongly correlated with Hubble type, and nearby systems
with $n \lesssim 2$ are generally disks, although low-mass early-type galaxies and dwarf spheroidals also mostly have low S\'ersic indices \citep[e.g.][]{Graham2003}. Repeating the K-S tests already described, but now using histograms of S\'ersic index instead of ellipticity, provides us with a different picture of the structural composition of the high-redshift compact population. As shown in Figure~2, the distribution of S\'ersic indices of local early-type galaxies is clearly inconsistent with that of high-redshift compact objects (see left panel Figure~\ref{fig_sersic}). The $p$-value returned by the K-S test is $\sim 10^{-15}$,  reinforcing the visual impression that the probability of these samples being drawn from the same underlying distribution is negligible. Interestingly, the K-S test does return a much larger $p$-value (60\%) when we compare the S\'ersic index distribution of the local massive disk-dominated galaxy sample with the high-redshift one, though only for the most massive $(M_* > 10^{11}M_\odot)$ systems (right panel of Figure~\ref{fig_sersic}). Selecting only quiescent (EW[OII] $<$ 5\AA) local galaxies does again not alter our conclusions: for local massive quiescent early-types $p \sim 10^{-14}$, although for local massive disk galaxies $p=33\%$. 

S\'ersic indices are challenging to measure in high-redshift galaxies, as possible outer wings of surface brightness profiles may be too faint to observe, biasing the observed S\'ersic indices towards lower values. As a consequence, the systematic errors on these measurements could be significant and influence our conclusions above. We therefore investigated the effect that measurement errors in the S\'ersic index could have on our results with Monte Carlo simulations. The errors in the Damjanov et al. sample were determined with Monte Carlo simulations which incorporated all the systematic and random errors that they were able to identify \citep[for details see][]{Damjanov2009}, exchanges in background \citep{VanDokkum2008} or GALFIT statistical errors \citep{Daddi2005}. We do not have errors for the S\'ersic indices presented in \citet{VanderWel2011}, but adopt for these galaxies the median value of the errors in the Damjanov et al. sample, which is 0.5.  We run the K-S test 10,000 times on our sample, each time adding Gaussian noise to our observed S\'ersic measurements based on these errors.

For the K-S tests comparing the S\'ersic indices of the high-redshift sample with the highest mass ($M_* > 10^{11} M_\odot$) local early-type galaxies, we obtain $p \sim 10^{-12}$. Even if we place all S\'ersic indices at their one-sigma upperlimit, we find $p \sim 10^{-11}$. We therefore conclude that the missmatch between S\'ersic indices of our compact high-redshift and local early-type sample is not a result of possible underestimation of the measured S\'ersic indices at high redshift.

\section{Discussion}

\begin{figure*}[tb]
\begin{center}
\begin{tabular}{ccc}
\includegraphics[width=5.2cm,trim=1.3cm 0cm 0cm 0cm]{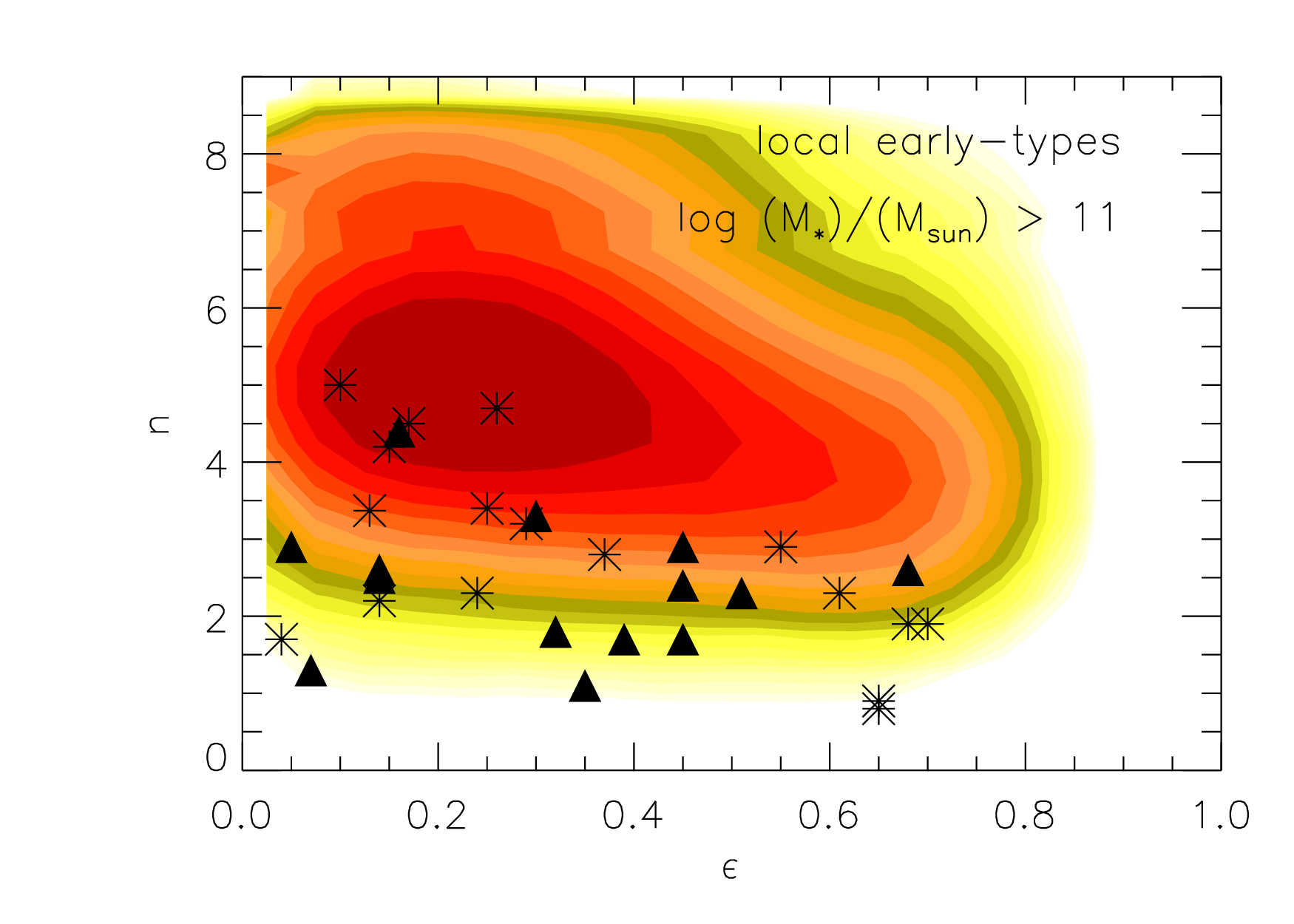} &
\includegraphics[width=5.2cm,trim=1.3cm 0cm 0cm 0cm]{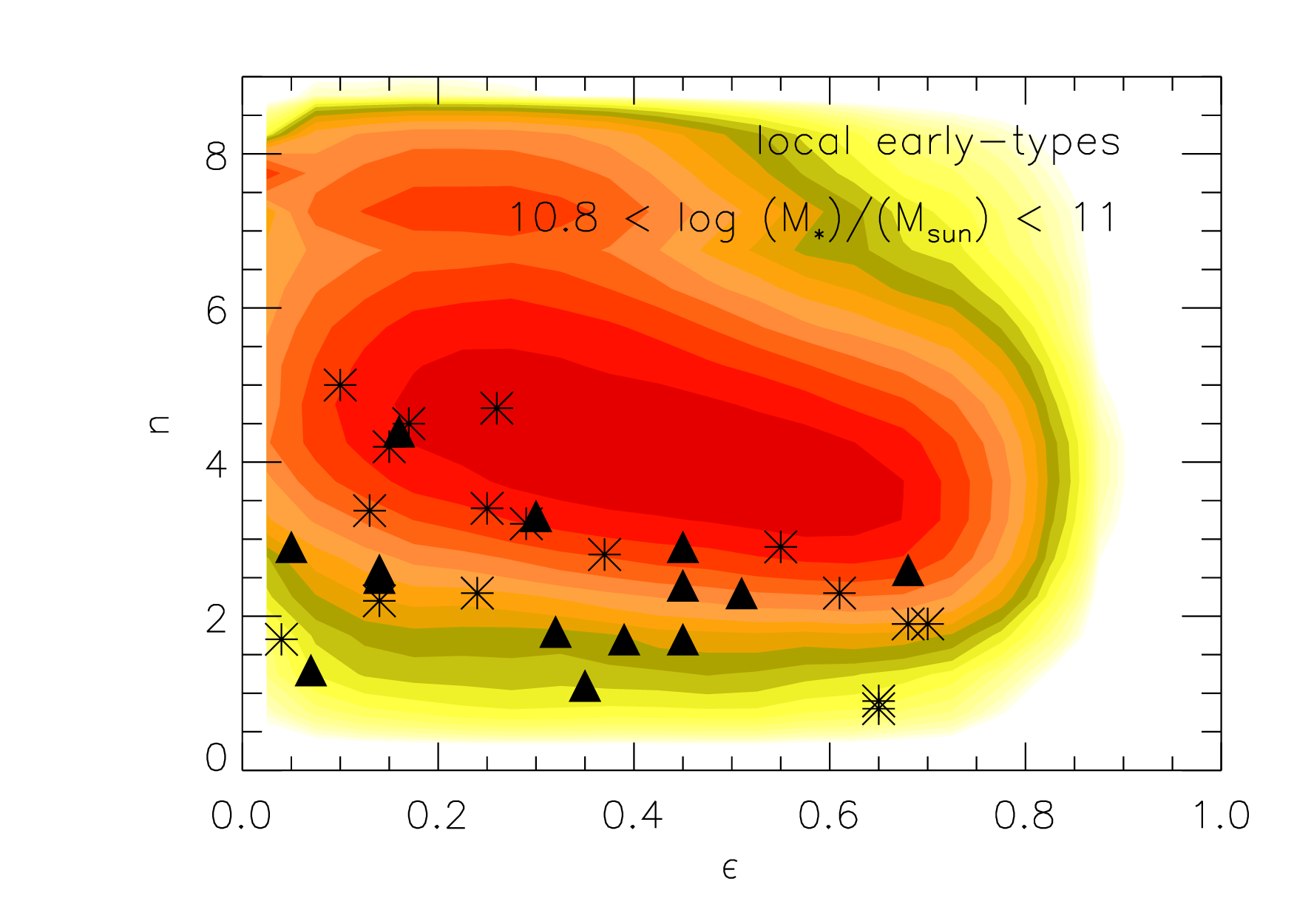} &
\includegraphics[width=5.2cm,trim=1.3cm 0cm 0cm 0cm]{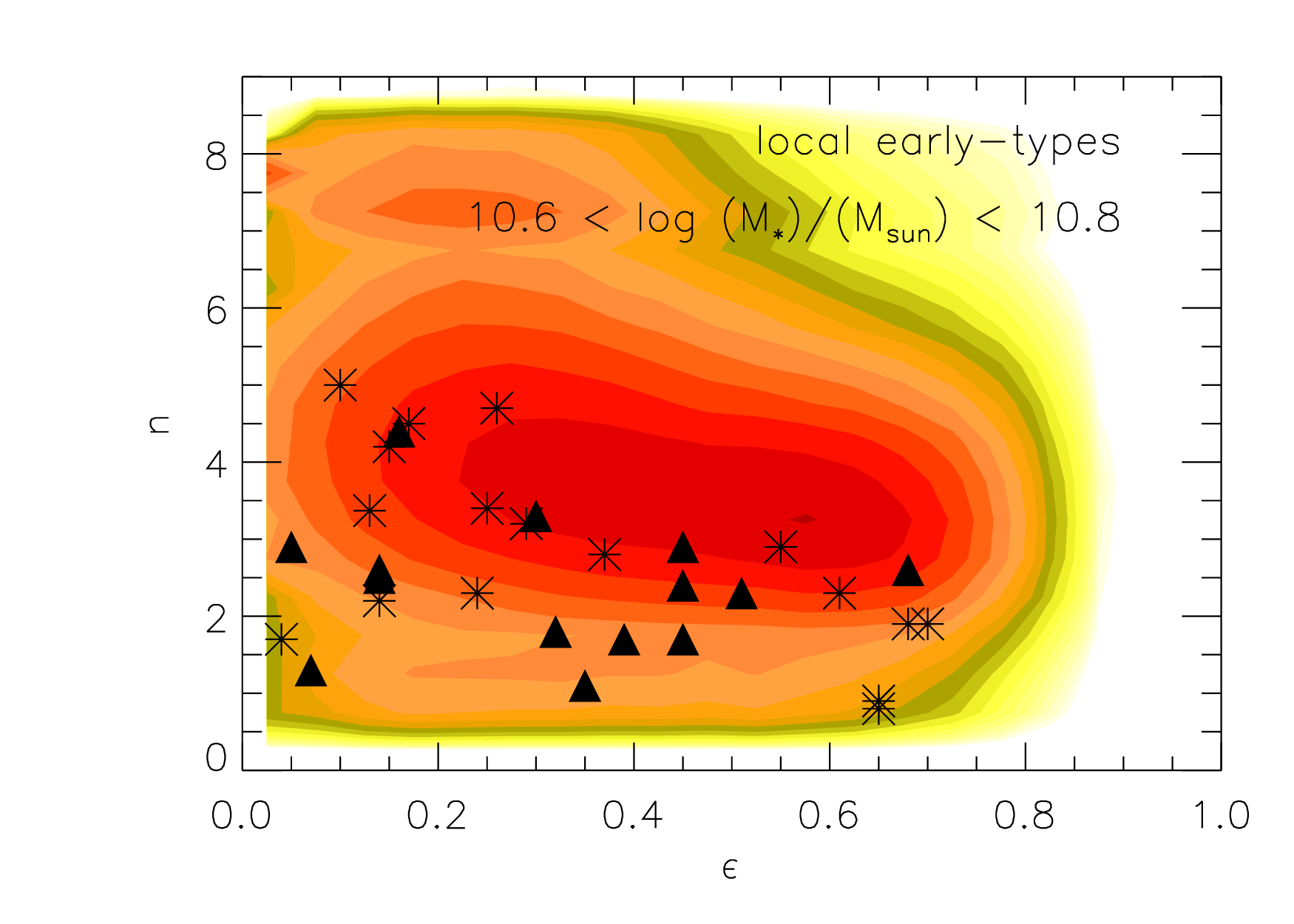} \\
\includegraphics[width=5.2cm,trim=1.3cm 0cm 0cm 0cm]{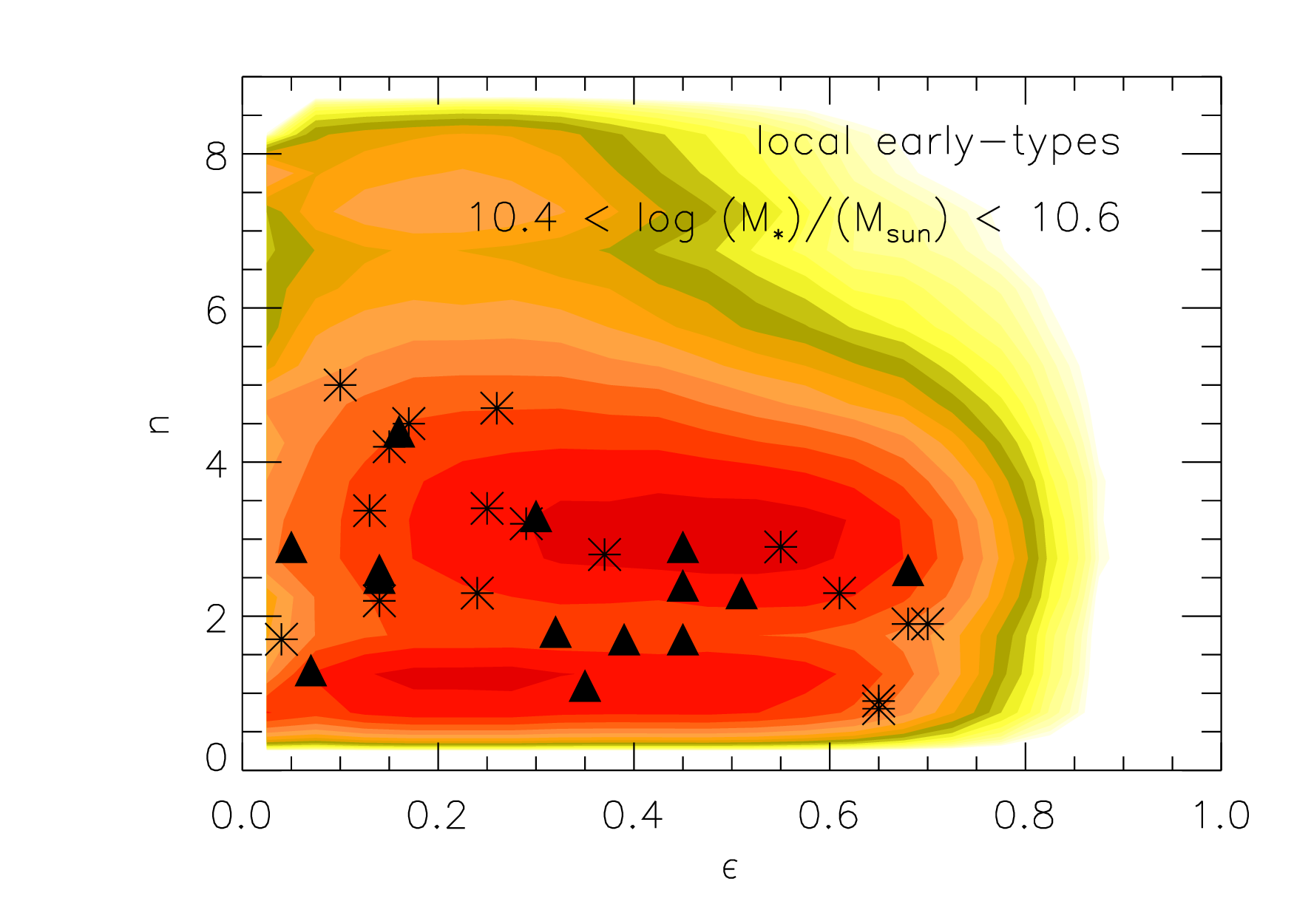} &
\includegraphics[width=5.2cm,trim=1.3cm 0cm 0cm 0cm]{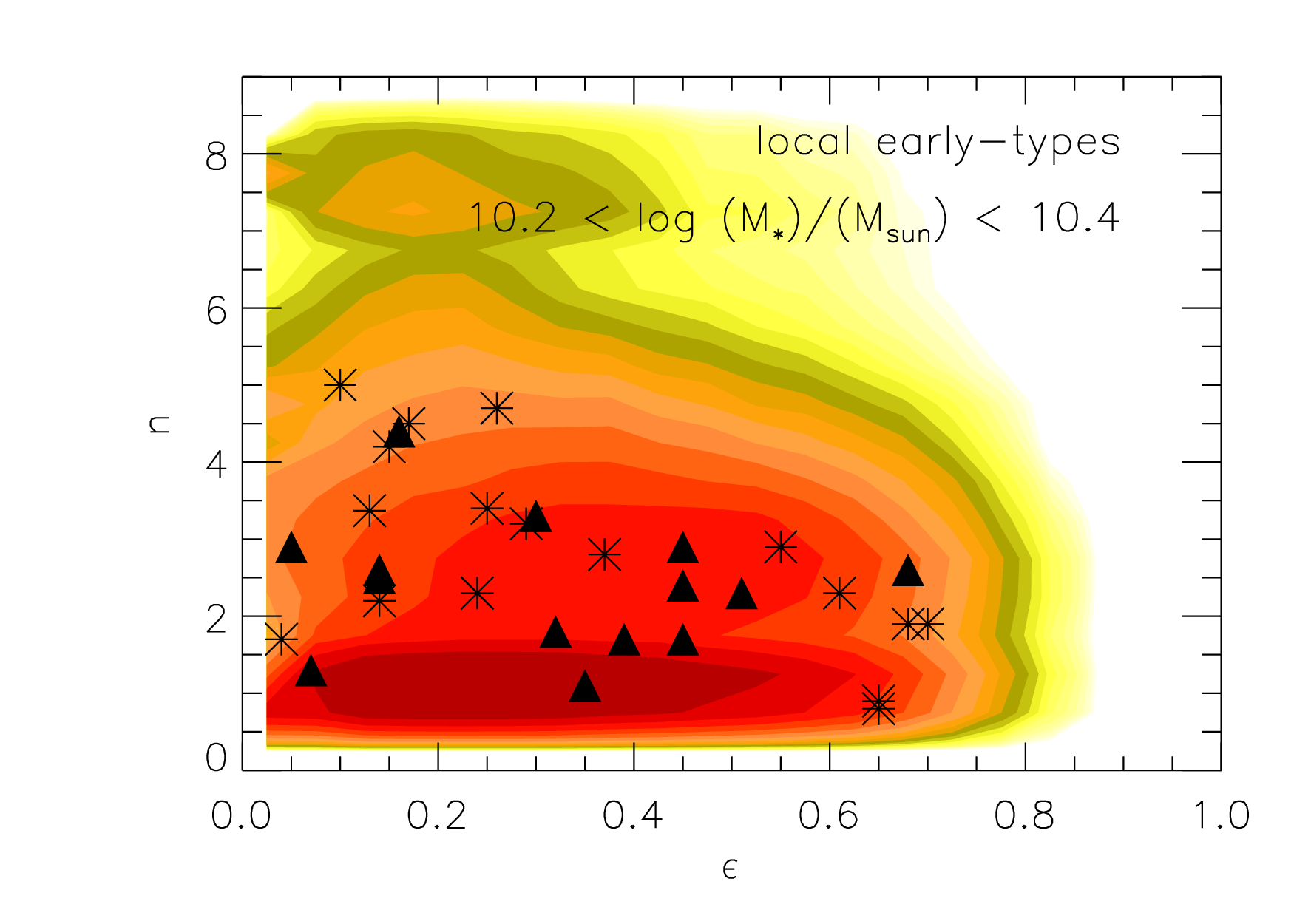} &
\includegraphics[width=5.2cm,trim=1.3cm 0cm 0cm 0cm]{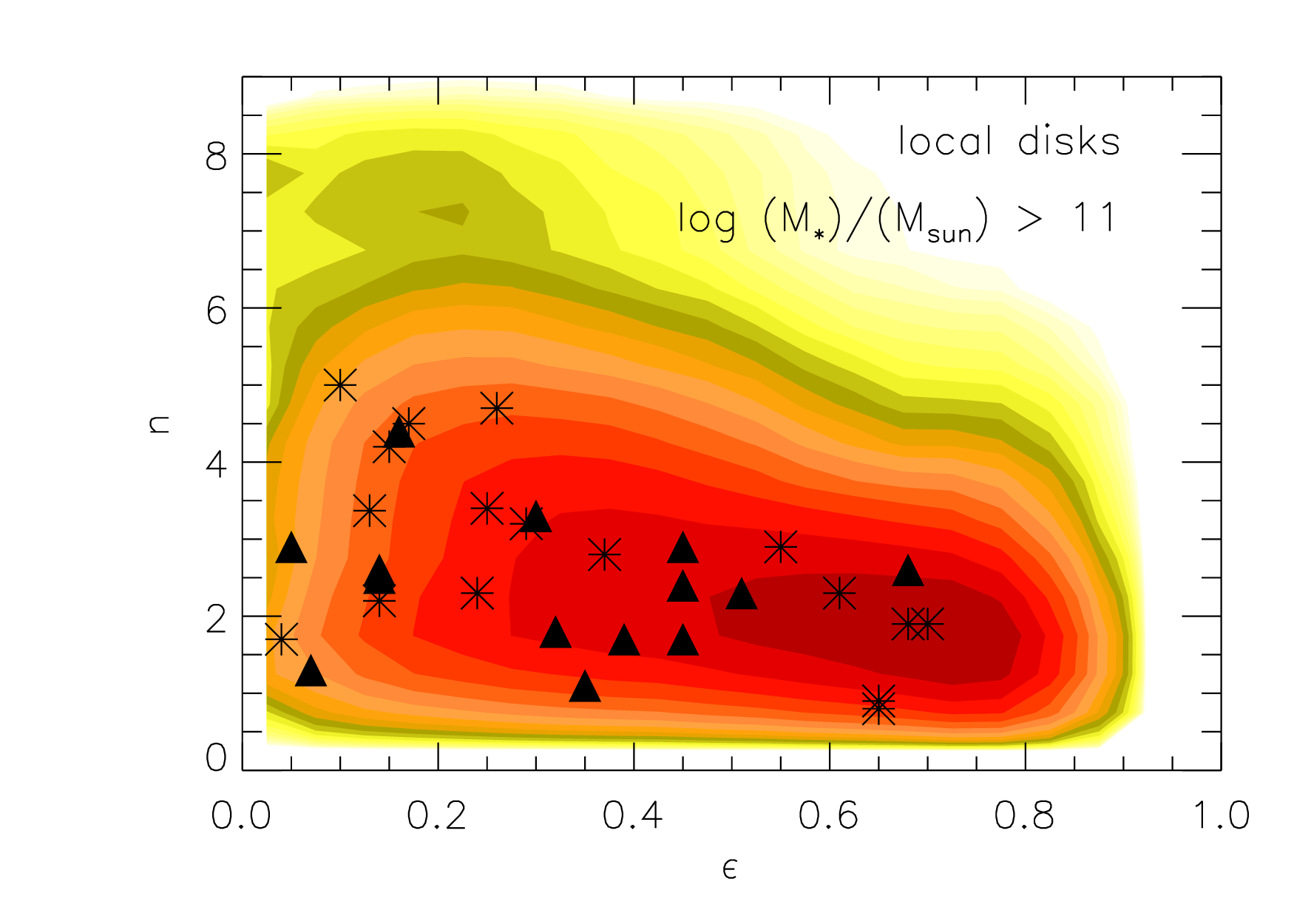} 
\end{tabular}
\end{center}
\caption{Comparing elliptiticity and S\'ersic index distributions from local {\it early-type} galaxies (contours) to those of the high-redshift massive compact galaxies of van der Wel et al. (2011, black filled triangles) and Damjanov et al. (2011, black stars). Contours are normalized and smoothed, increasing in number density from yellow to red in logarithmic steps. Different panels show different mass ranges for the local early-type galaxy sample, denoted in the upper right corner or each panel. The last panel (lower right) shows the distribution for local massive {\it disk-dominated} galaxies.}
\label{fig_eps_n}
\end{figure*}

Compact massive quiescent galaxies at $z\sim2$ have an ellipticity distribution consistent with that of local massive early-type galaxies, but their S\'ersic index distribution seems to rule out the possibility these high-$z$ objects are all early-type galaxies.  The S\'ersic index distribution of the high-redshift galaxies is fairly similar to that of local disk populations, though the ellipticity distributions are very different. What are we to make of this conundrum? One possibility is that the population of high-redshift compact massive galaxies is a mixture of bulge and disk dominated galaxies. Such a model might be made to work, but inspection of Figures~1 and 2 suggests that it would not be very straightforward to mix ellipticity and S\'ersic indices for early-type and disk-galaxies in such a way as to have bulge-dominated galaxies dominate one property, while disk-dominated galaxies dominate the other.  Another possible problem with this model is that it would require galaxies, that have different structural properties, to all change in size with cosmic epoch, which seems hard to credit unless the growth mechanism is somehow extrinsic to the galaxies themselves \citep[a hypothesis briefly discussed in][]{Damjanov2011}.

An alternative, and in some ways more natural explanation for the discrepancy between combined structural properties of our samples at $z\sim2$ and $z\sim0$ might simply be that the high-redshift galaxy population has unique properties, that are not present in the local galaxy population because local galaxies are poor analogs to the high-$z$ systems. In other words, the high-redshift compact galaxies may truly be a class on their own. 

In Figure~\ref{fig_eps_n} we explore these alternatives a little further by plotting ellipticity versus S\'ersic index for our high-redshift sample on top of contours showing the distribution of these same parameters in
local galaxies. Different panels in this figure correspond to
different mass bins. Table~\ref{tab} summarizes the results for the various K-S tests that we ran on all these samples. None of the local galaxy samples provide a good fit to the observed distributions at high redshift, and the fact that also the lower-mass early-type galaxy samples fail to do so rules out the possibility that our results are due to errors in assigning stellar masses to the galaxies (perhaps as the result of an evolving IMF). Interestingly, a simple likelihood analysis\footnote{ We calculate the likelihood L as $\log L = \sum_{i=1}^{N} P(\epsilon_i, n_i) \textrm{d}\epsilon \textrm{d}n$, for our sample of $N=31$ high-$z$ galaxies, with $P(\epsilon, n)$ defined by the normalized distribution of the local galaxy sample.} shows that the disk-dominated sample (most-right bottom panel) has a higher likelihood than the early-type local samples, when taking both the ellipticity and S\'ersic distributions into account simultaneously. This likelihood analysis however only tells us that one distribution is a better fit than the other, but not that this fit is necessarily a good one.

Figure~\ref{fig_eps_n} highlights another important point. Although no local bivariate ellipticity-S\'ersic index distribution is a good fit to the high-$z$ data, there certainly does exist a strong overlap in parameter space between the high-$z$ population and the local sample. This fact, together with inspection of the images presented in \citet{VanderWel2011}, but also \citet{Daddi2005,Longhetti2007,Toft2007,VanDokkum2008,Cimatti2008,Damjanov2009,Ryan2010,Szomoru2011}, makes it seem fairly clear that {\em some} of these objects are indeed disks. 
Recent results of \citet{Weinzirl2011} shown that although the majority of {\it all} massive galaxies at $2\lesssim z\lesssim3$ show possible disk-like structures with $n\lesssim2$ \citep[see also][]{Buitrago2011}, most ($\sim72\%$) of these objects are extended ($R_e>2$~kpc). On the other hand, half of the population of compact (predominantly quiescent) objects do not show any disk-like features, and for the highest mass bin ($M_* >10^{11}M_\odot$) only $\sim38\%$ has $n\leq2$.

Our results lead to a prediction that can be used to test the formation mechanism(s) of massive galaxies. Major gas-rich mergers are believed to produce compact spheroidal remnants with light profiles characterized by high S\'ersic indices \citep[e.g.,][]{Wuyts2010,Bournaud2011}. Additional {\it gradual and slow} gas accretion is required to build up a disk component \citep{Weinzirl2011,VanderWel2011}. One intriguing possibility is that compact quiescent galaxies at high-redshift consist of thick disks, possibly formed by star-forming turbulent disks at $z > 2$ \citep[e.g.][]{Elmegreen2006,Bournaud2009}. The thick disk would then be responsible for the mismatch with the ellipticity distributions of the local disk-dominated population, with the disk thickness driving the lack of objects with extreme ellipticities $(\epsilon > 0.7)$ compared to modern spirals, whose light is dominated by thin disks. 

It is worth emphasizing that if quiescent populations at $z=0$ and $z\sim2$ have very different galaxy light profiles, which is one possible interpretation of our data,  then the subsequent evolution of compact quiescent galaxies bearing disks would need to involve a drastic change in their S\'ersic indices. In a scenario in which their size evolution is driven mainly by a series of minor mergers, the S\'ersic indices of compact disk galaxies can change by adding a small amount of low density stellar mass to their outskirts \citep[e.g.,][]{Bournaud2007}. At this point it is not clear whether that change would be dramatic enough to turn compact disks into systems resembling the spheroids found locally \citep{Weinzirl2011}.

\begin{table*}[tb]
\begin{center}
\begin{tabular}{l|l|l|l|l}
Type & Mass Range & N & ellipticity & S\'ersic index  \\
     &    ($\log M_\odot$) & &   $p$-value&  $p$-value \\
\hline
early-type &  11.0 -  & 39,996 &  0.82   & 0.00  \\
  & 10.8 -   & 63,026 & 0.62  & 0.00 \\
  & 10.8 - 11.0  & 23,020 & 0.31 & 0.00 \\
 & 10.6 - 10.8  & 21,296 & 0.28 & 0.00 \\
 & 10.4 - 10.6  & 15,053 & 0.45 & 0.07 \\
 & 10.2 - 10.4  & 10,297 & 0.64 & 0.00 \\
\hline
disk &  11.0 -  & 30,098 &  0.05   & 0.60  \\
     &  10.8 -  & 56,763 &  0.03   & 0.07  \\
\end{tabular}
\end{center}
\caption{Sample sizes and Kolmogorov-Smirnov test $p$-values (probability that two samples were drawn from the same underlying distribution) for comparisons in ellipticity and S\'ersic index between the local and high-redshift sample, for several mass bins. The local early-type sample is chosen such that $B/T > 0.5$ and smoothness parameter $S2 \leq 0.075$; the local disk sample is selected such that $B/T < 0.5$, with no additional constraints on smoothness. See text (\S2) for more details.}
\label{tab}
\end{table*}

\section{Conclusion}\label{Con}

We have compared the observed ellipticity distribution of massive, compact, quiescent galaxies at high redshift ($1.5 < z < 2.5$) to those of local ($z < 0.1$) early-type galaxies, and conclude that the two distributions are statistically indistinguishable. In addition, we show that the ellipticity distribution of our high-$z$ sample is inconsistent with that of a local massive disk-dominated sample. On the other hand, the S\'ersic index distributions of compact high-$z$ galaxies and local early-type galaxies are not consistent, and the S\'ersic index distribution of local disk galaxy samples provides a better match to the high-$z$ data. We  conclude that either (a) high-$z$ galaxies are a composite population of disks and bulges (which presents the troublesome possibility that galaxy sizes are growing for all structural types); or else, (b) that the shapes and light profiles of high-redshift massive compact galaxies are unlike those of any local galaxy sample, and their structure changes over time. In the latter case, these objects constitute a new class of galaxies on their own.

\acknowledgements
 We thank the referee for his/her constructive comments, which helped to improve the presentation of this paper. This work makes use of the Sloan Digital Sky Survey (www.sdss.org).


\end{document}